\documentclass{aa}
\usepackage{graphicx}
\usepackage{rotating}
\usepackage{afterpage}
\usepackage[sort&compress]{natbib}
\bibpunct{(}{)}{;}{a}{}{,}
\newcommand{\kms}{km\thinspace s$^{-1}$}

\newcommand{\ks}{{\it K$_{\rm S}$}}

\newcommand{\Av}{{\it A$_{\rm V}$}}

\begin{document}
   \title{86 GHz SiO maser survey of late-type stars in the Inner Galaxy 
   \thanks{ This is paper no. 21 in a refereed journal based on
   data from the ISOGAL project. \newline Based on observations with
   ISO, an ESA project with instruments funded by ESA member states
   and with the participations of ISAS and NASA.\newline Based on
   observations collected at the European Southern Observatory, La
   Silla Chile. \newline Based on observations carried out with the
   IRAM 30-m telescope located at Pico Veleta. IRAM is supported by
   INSU/CNRS (France), MPG (Germany) and IGN (Spain).}  }

   \subtitle{II. Infrared photometry of the SiO Target Stars}
   \author{M.\ Messineo
          \inst{1}
          \and
          H.\ J.\ Habing
          \inst{1}
          \and
          K.\ M.\ Menten
          \inst{2}
          \and
          A.\ Omont
          \inst{3}
          \and
          L.\ O.\ Sjouwerman
          \inst{4}
	      }

   \offprints{M. Messineo}

   \institute{Leiden Observatory, P.O. Box 9513, 2300 RA Leiden, the Netherlands\\
              \email{messineo@strw.leidenuniv.nl; habing@strw.leidenuniv.nl}
         \and
             Max-Planck-Institut f\"ur Radioastronomie, Auf dem H\"ugel 69, D-53121 Bonn, Germany\\
             \email{menten@mpifr-bonn.mpg.de}
         \and Institut d'Astrophysique de Paris, CNRS \& Universit\'e 
Paris 6, 98bis Bd Arago, F 75014 Paris, France\\
           \email{omont@iap.fr}
         \and
             National Radio Astronomy Observatory, P.O. Box 0, Socorro NM 87801, USA\\
             \email{lsjouwerman@aoc.nrao.edu}
             }

   \date{Received October 09, 2003; accepted December 19, 2003.}

   \abstract{We present a compilation and study of DENIS, 2MASS,
ISOGAL, MSX and IRAS 1--25$\mu$m photometry for a sample of 441
late-type stars in the inner Galaxy, which we previously searched for
86 GHz SiO maser emission \citep{messineo02}.  The comparison of the
DENIS and 2MASS $J$ and \ks\ magnitudes shows that most of the SiO
targets are indeed variable stars.  The MSX colours and the IRAS
$[12]-[25]$ colour of our SiO targets are consistent with those of
Mira type stars with dust silicate feature at 9.7 $\mu$m feature in
emission, indicating only a moderate mass-loss rate.   }

   \maketitle
%
\section{Introduction}
Stars of intermediate mass, $1<M_*<6\, M_\odot$, enter a phase of
intense mass loss when they reach the Asymptotic Giant Branch
(AGB). As a consequence, they are surrounded with a dense envelope of
dust and molecular gas.  Due to the low effective temperature and the
dust thermal emission, AGB stars are bright at infrared wavelengths
and can be detected even towards highly obscured regions. Furthermore,
the maser emission from their envelopes is strong enough to be
detected throughout the Galaxy, and radio spectroscopic observations
can provide the stellar line-of-sight velocities to within a few \kms\
\citep[e.g.][]{habing96}. AGB stars thus permit a study of Galactic
kinematics, structure, and mass-distribution.

To understand the Galactic structure and kinematics it is important to
combine the kinematic information and the stellar properties, e.g.
luminosities, which can provide a distance estimate.  Good photometry
on infrared point sources toward the inner Galaxy is now available
from large surveys such as DENIS \citep{epchtein94}, 2MASS
\citep{2massES}, ISOGAL \citep{omont03,schuller03} and MSX
\citep{egan99,price01}.  Since the high extinction toward the inner
Galaxy precludes studies at optical wavelengths, these infrared data
permit a unique view of its stellar population.  The combination of
near- and mid-infrared photometry enable us to examine the nature of
the stars, i.e. to derive their luminosities, mass-loss rates, and to
discriminate againts foreground stars.

To improve the line-of-sight velocity statistics, we conducted 86 GHz
$v = 1, J = 2 \rightarrow 1$ SiO maser line observations of 441
late-type stars in the inner Galaxy ($30^\circ < l < -4^\circ$, $|b| <
1$) with the IRAM 30-m telescope \citep[][hereafter Paper\,I]
{messineo02}.  \defcitealias{messineo02}{Paper\,I} This paper is part
of a series devoted to characterise the properties, i.e. mass-loss
rates and luminosities, of the 441 sources previously targeted to
search for 86 GHz SiO maser emission \citepalias{messineo02}.

Here (Paper\,II) we present the available near- and mid-infrared
photometry of the targeted sources (``SiO targets'' hereafter). In
another paper \citep[][hereafter Paper\,III] {messineo03_3}
\defcitealias{messineo03_3}{Paper\,III} we deal with extinction
correction and finally in the last paper \citep[][hereafter
Paper\,IV] {messineo03_4} \defcitealias{messineo03_4}{Paper\,IV} we
compute and analyse the luminosities of the SiO targets.

\begin{figure*}[ht!]
\includegraphics[width=\textwidth]{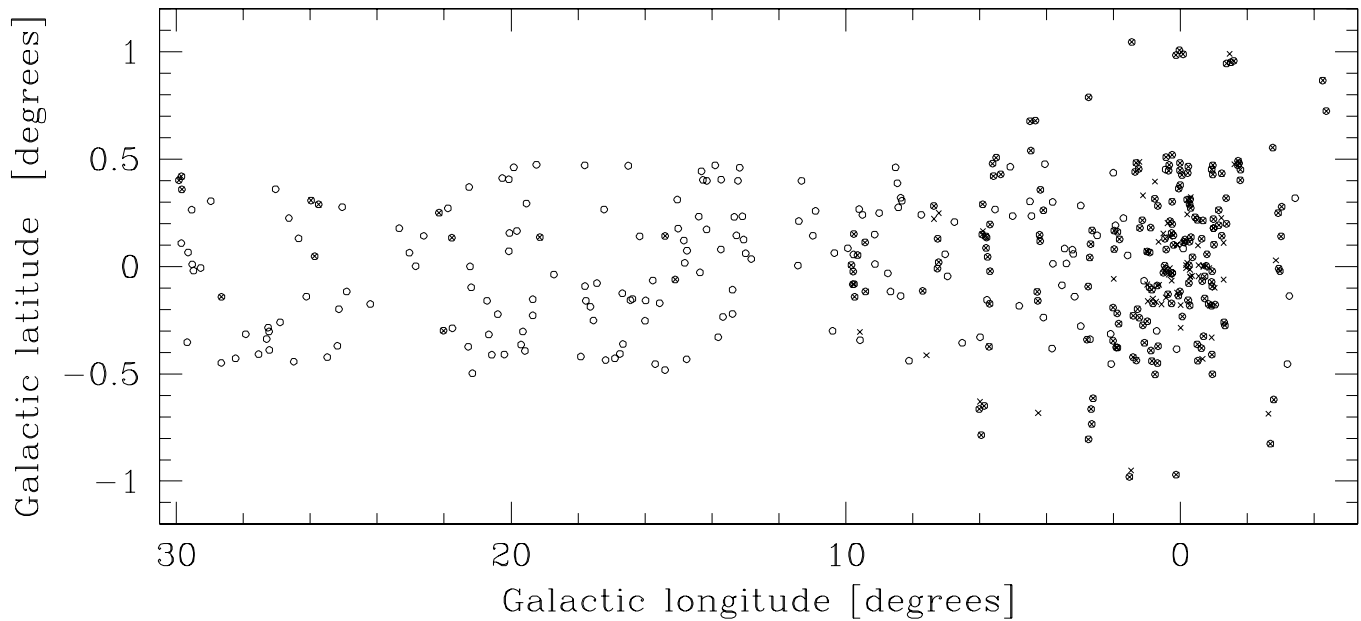}
\caption{\label{fig:lb.ps} Location of the 441 SiO targets in
Galactic coordinates.  The 379 MSX counterparts are shown as open
circles, the 267 ISOGAL counterparts as crosses. Overlap of MSX and
ISOGAL sources resemple filled symbols. Four points fall outside the
figure.}
\end{figure*}

The spatial distribution of the 441 targets is shown in Fig.\
\ref{fig:lb.ps}.  These SiO targets are divided into two subsamples:
253 sources were selected from the ISOGAL catalogue, (``the ISOGAL
sample''), and 188 sources from the MSX catalogue, (``the MSX
sample''). The ISOGAL and MSX samples are examined to test
whether both are drawn from the same ``parent population''.
Brightness variability is studied by a comparison of the DENIS and
2MASS photometry.

The structure of the paper is as follows: in Sect.\ \ref{database} we
identify our SiO targets in various infrared catalogues and collect
their magnitudes finding for many stars up to fourteen different
measurements. In Sect.\ \ref{first} we compare the statistical
differences between our ISOGAL and MSX samples. In Sect.\
\ref{remarks} we summarise additional information found with SIMBAD,
e.g.\ variability and other types of masers, and we derive the
probability of association between the radio maser and the infrared
counterpart.  The brightness variability of the stars is discussed in
Sect.\ \ref{variability}.  In Sects.\ \ref{irascolours} and
\ref{msxcolours} we analyse the mid-infrared colours of the stars and
compare them with those of OH/IR stars.  The main conclusions are
summarised in Sect.\ \ref{conclusion}.

The individual source numbers (e.g \#99) are taken from Table 2 (86
GHz SiO maser detections) and Table 3 (non-detections) in
\citetalias{messineo02} unless otherwise indicated.  The SiO maser
emission in this paper generally refers to the 86 GHz ($v=1,
J=2\rightarrow 1$) SiO maser only and not to the 43 GHz ($J=1-0$) SiO
masers. Velocities in this paper refer to line-of-sight velocities
with respect to the Local Standard of Rest.

\section{Identification of the SiO targets in various infrared catalogues}
\label{database}

We cross identified all SiO targets, whether taken from the ISOGAL or
from the MSX database, with all infrared catalogues available to us,
the DENIS, 2MASS, ISOGAL, MSX and IRAS survey catalogues.  The results
are summarised in Table \ref{table:counterparts}.  In the following,
we briefly recall the criteria used for the selection of the ISOGAL
and the MSX samples and describe the modality of the
cross-correlations. More details on the selection criteria can be
found in \citetalias{messineo02}.

\begin{table*}[bht!]
\caption{\label{table:counterparts} Number of counterparts of our SiO targets}
\begin{tabular}{l|rrr|rrr|ccc}
\hline
\hline
{\rm~~} &\multicolumn{3}{c}{\rm 2MASS} & \multicolumn{3}{c}{\rm DENIS} & {\rm ISOGAL} & {\rm MSX} & {\rm IRAS}\\
{\rm~~} &\multicolumn{3}{c}{\rm all sky data release } & \multicolumn{3}{c}{\rm bulge PSC } & {\rm PSC 1.0} & {\rm PSC 1.2} & {\rm PSC 2.0}\\
{\rm~~} &\multicolumn{3}{c}{\rm \citep{2massES}} & \multicolumn{3}{c}{\rm \citep{simon03}} & {\rm \citep{schuller03}} & {\rm \citep{egan99}} & {\rm \citep{iras}}\\
{\rm}&{\rm J}&{\rm H}&{\rm $K_S$}&{\rm I}&{\rm J}&{\rm K$_S$}&&&\\
\hline
{\rm ISO~~~~}(253) &     252&252&252           &53&217&253                     &   253        & 191       &   ~43 \\ 
{\rm MSX~~}(188) &     187&187&187           &~42&149&188                     &    ~14        & 188       &   122 \\
\hline
{\rm Total}~~(441) &     $^{\mathrm + }$439&$^{\mathrm ++ }$439&439           &95&$^{\mathrm * }$366 & $^{\mathrm **}$441&267& 379 &   165 \\
\end{tabular}
\begin{list}{}{}
\item[]
$^{\mathrm{+}}$ 
~~72 are upper limits
\hspace{0.5 cm} 
$^{\mathrm{*}}$ 
~~16 have magnitudes above saturation limits
\item[]
$^{\mathrm{++}}$
13 are upper limits
\hspace{0.5cm} 
$^{\mathrm{**}}$
104 have magnitudes above saturation limits
\end{list}
\end{table*}

\subsection{ISOGAL data}\label{isogal}
ISOGAL \citep{omont03} is a 7 and 15 $\mu$m survey taken with ISOCAM
\citep{cesarsky96} aboard the Infrared Space Observatory (ISO)
satellite \citep{kessler96}. The 16 deg$^2$ survey consists of
selected sub-fields along the Galactic plane, mostly concentrating on
the Galactic Centre.  With a sensitivity approaching 10 mJy (two
orders of magnitude deeper than IRAS and one order of magnitude deeper
than MSX) and a resolution of 3-6\arcsec\, ISOGAL has detected over
100,000 objects.  The ISOCAM data have been correlated with the DENIS
near-infrared {\it I, J} and \ks\ band data to produce the five band
ISOGAL-DENIS point source catalogue (ID-PSC) \citep{schuller03}. The
ID-PSC reports the photometric data in magnitudes.  The astrometric
accuracy of the ID-PSC is determined by the present DENIS astrometric
accuracy better than 0.5\arcsec.

Sources were selected from a preliminary version of the ISOGAL
catalogue by their extinction-corrected 15 $\mu$m magnitude, $[15]_0$,
and their (\ks$_0-[15]_0)$ and ($[7]_0-[15]_0$) colours approximately
corrected for extinction \citepalias[Sect.\ 7.3.2][]{messineo02}. The
brightest 15 $\mu$m sources, [15]$_0<1.0$, and those with
([7]$_0-[15]_0) < 0.7$ and with (\ks$_0-[15]_0) < 1.95$ were excluded
since they are likely to be foreground stars or non-AGB stars or AGB
stars with very small mass-loss.  Further, sources with $[15]_0 >$ 3.4
were excluded since they are likely to show SiO maser emission fainter
than our detection limit of 0.2 Jy.  Sources with ($[7]_0-[15]_0) >
2.3 $ were excluded since they are likely to be compact H\ion{II}\
regions or other young stellar objects or planetary nebulae. Those
with (\ks$_0-[15]_0) > 4.85$ were excluded because they are likely to
be OH/IR stars with a high mass-loss rate or young stellar objects.
Moreover, known OH/IR stars were discarded as the kinematic data are
already known.

\subsection{MSX data}

The Midcourse Space Experiment (MSX) is a survey at five mid-IR bands
ranging from 4.3 $\mu$m [$B1$ band] to 21.4 $\mu$m [$E$ band], with a
sensitivity of 0.1 Jy in $A$ band (8.28 $\mu$m) and a spatial
resolution of 18.3\arcsec\ \citep{price01}.  The survey covers the
Galactic plane to $\pm 5 ^\circ $ latitude.  Version 1.2 of the
MSX-PSC \citep{egan99} lists more than 300,000 point sources with an
rms astrometric accuracy of $\sim2$\arcsec. The MSX catalogue gives
the source flux density, $F$, in Jy.  Magnitudes are obtained adopting
the following zero-points: 58.49 Jy in $A$ (8.26 $\mu$m) band, 26.51
Jy in $C$ (12.12 $\mu$m) band, 18.29 Jy in $D$ (14.65 $\mu$m) band and
8.80 Jy in $E$ (21.41 $\mu$m) band \citep{egan99}.

For the MSX source selection we used flux densities in the $A$ and $D$
bands which have wavelength ranges roughly similar to the ISOGAL 7 and
15 $\mu$m bands.  We selected those non-confused, good-quality sources
in $A$ and $D$ band (flag $> 3$), which show variability in the $A$
band.  We avoided the reddest stars, $F_D/F_A > 2.3$ ($A-D > 2.2 $
mag). We also avoided the bluest and most luminous stars with $F_D/F_A
< 0.6$ ($A-D < 0.75$ mag) and $F_D > 6$ Jy ($D < 1.2$ mag) since they
are likely to be foreground stars or supergiants
\citep{schuller02,schultheis03}.  Furthermore, following the
classification of \citet{kwok97} of IRAS sources with low-resolution
spectra, we used the $C$ to $E$ band ratio to discard very red
($F_E/F_C > 1.4$, $C-E > 1.55$ mag) sources, which are likely to be
young stellar objects or OH/IR stars with thick envelopes.  Known
OH/IR stars were excluded.

\subsection{ISOGAL-MSX cross-identifications}\label{msxiso-x}
\begin{figure}[t]
\resizebox{\hsize}{!}{\includegraphics{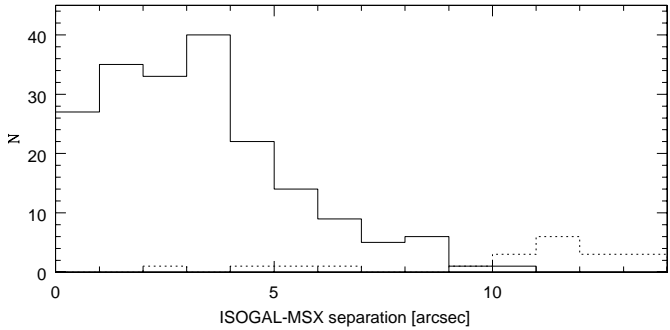}}
\caption{\label{fig:isomsxdist.ps} Associations between our SiO ISOGAL
targets and the MSX catalogues.  The distribution of the angular
separations (\arcsec) between the ISOGAL and the MSX positions
(continuum line). For comparison, the dashed line is the total
distribution of the chance associations, found  by performing
five shifting experiments.}
\end{figure}

More than half (253) of our sample of SiO targets were selected from
the ISOGAL survey, indeed from a preliminary version of the ID-PSC
\citepalias{messineo02}. They are all in an area covered by the MSX
survey.
For each of our 253 ID-PSC selected SiO targets we searched for the
closest associated source in the MSX-PSC within 15\arcsec\ from the
SiO target position. We found counterparts for 191 of them, 190 of
which were unique.  The distribution of the angular separations is
shown in Fig.\ \ref{fig:isomsxdist.ps}; the mean and median angular
separation are 3.3\arcsec\ and 3.0\arcsec\ with a standard deviation
of 2.1\arcsec, the maximum separation is 10.4\arcsec.  To estimate the
likely number of chance associations, we repeated the
cross-correlation after shifting all ID-PSC positions between
30\arcsec\ and 100\arcsec\, finding that the probability of spurious
associations within 11\arcsec\ is less than 1\%.  The 62 ISOGAL
sources without a counterpart in the MSX-PSC are mostly (55)
concentrated in the central 3 degrees from the Galactic Centre.  Their
7 $\mu$m ISOGAL fluxes range from 0.4 Jy to 1.5 Jy, whereas for the
191 associated sources, the flux is on average 1.6 Jy with a standard
deviation of 1.3 Jy. Thus, the ISOGAL/non-MSX sources are likely to
have been excluded from the MSX-PSC because of confusion due to the
high stellar density in the inner Galaxy. The fraction of our ISOGAL
stars without MSX identifications is $\sim$11\% for those with 7
$\mu$m magnitude $[7] < 4.5$ ($F_7>1.3$ Jy).

Out of the 188 SiO targets selected from the MSX catalogue, only 14
are located in an ISOGAL field, and we found ID-PSC identifications
for all of those within 5\arcsec\ from the MSX positions.  For the
remaining 174 sources we searched the DENIS PSC.

\begin{figure}[t]
\resizebox{\hsize}{!}{\includegraphics{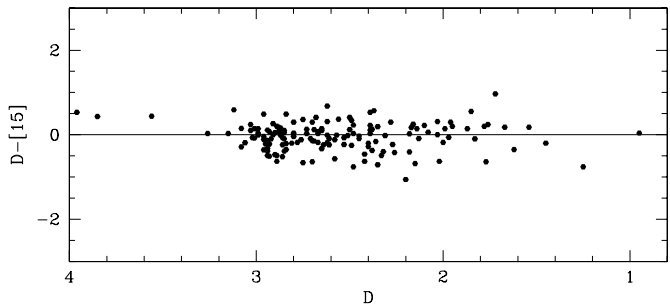}}
\caption{\label{fig:diff15.ps} Difference between the MSX band $D$
magnitude, $D$, and the ISOCAM 15 $\mu$m magnitudes, [15], versus the
$D$. Magnitude zero point in $D$ band was taken from \citet{egan99}.
The continuous line is $D-[15]=0.0$ mag}
\end{figure}

As the ISOCAM 15 $\mu$m and MSX $D$ filters are similar, we compared
the ISOCAM 15 $\mu$m magnitudes, [15], and the MSX $D$ band
magnitudes, $D$, of the 154 sources detected at 15 $\mu$m in both
surveys and found good agreement (Fig.\ \ref{fig:diff15.ps}); the
average difference $D-[15]$ is $-0.04$ magnitude and the standard
deviation is 0.3 mag, resulting from the combination ($\sim0.15$ mag)
of the photometric errors of both catalogues and from the possible
intrinsic source variability.

\subsection{DENIS data}
DENIS is a simultaneous $I$ (0.8 $\mu$m), $J$ (1.25 $\mu$m), and \ks\
(2.15 $\mu$m) band survey of the southern hemisphere using the ESO
1-meter telescope at La Silla \citep{epchtein94}.  The 3$\sigma$
detection limit in the respective bands is 0.05 mJy (19 mag), 0.5 mJy
(16 mag) and 2.5 mJy (13.5 mag), and the saturation limit is at
magnitudes 10, 7.5 and 6, respectively.  The absolute DENIS astrometry
is fixed using the USNO-A2.0 catalogue.  The current absolute
astrometric accuracy of DENIS is better than 0.5\arcsec\ (rms)
\citep[see Sect.\ 4.2 in][]{schuller03}.

\subsubsection{ISOGAL sample and DENIS identifications}
For the SiO targets selected from the ISOGAL survey, a proper DENIS
identification and photometry had already been provided
\citep{schuller03,simon03}.  
For each ISOGAL field, a sample of DENIS point sources from
the same area was extracted.  DENIS coordinates were
retained as the astrometric reference system and a polynomial
distortion correction was applied to the ISOGAL coordinates in order
to match as best as possible the ISOGAL and DENIS reference
coordinates.  In order to reduce the number of spurious associations,
only DENIS sources with a \ks\ detection were selected and a \ks\
upper limit was imposed according to field density \citep{schuller03}.
Among our 253 ISOGAL targets all have a DENIS \ks\ identification,
while only 86\% and 21\% have  $J$ and $I$ detections, respectively.

\subsubsection{MSX sample and DENIS identifications}
\label{msxden}
We searched for possible DENIS counterparts to the 188 SiO targets
which we had selected from the MSX catalogue.  For all sources from
the MSX sample we searched the provisional bulge DENIS PSC
\citep{schuller03,simon03} for the nearest \ks\ band object.  As
late-type stars are intrinsically red, only \ks\ band counterparts
were examined.  As saturated sources are usually given in the
catalogue with null magnitude, DENIS photometry from other
observations \citep{epchtein94} and the \ks\ images were used as
additional checks for saturated counterparts to the MSX SiO targets.
Fortunately, such DENIS data was available for all MSX sources that we
observed; however, a large number of bright sources saturated the
detector in the \ks\ band.  While all sources were detected in the
\ks\ band, the fraction of detections in the $J$ and $I$ band was 79\%
and 22\%, respectively.
\begin{figure}[!]
\begin{centering}
\resizebox{\hsize}{!}{\includegraphics{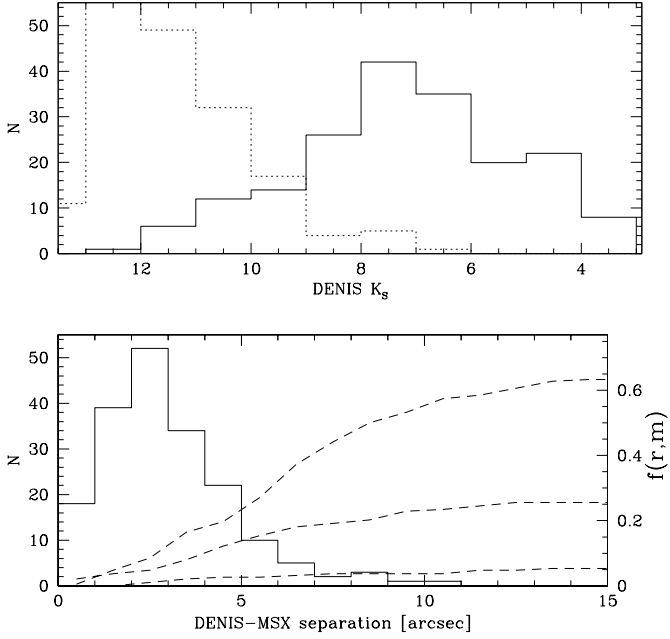}}
\caption{\label{fig:denismsx.ps} Associations of our SiO MSX targets
in the DENIS catalogue.  {\bf Top panel}: distribution of \ks\
magnitudes for the associations (continuum line) and for the chance
associations (dashed line). {\bf Lower panel}: The continuous line
shows the distribution of angular separations.  The dashed lines show
the normalised cumulative distributions, $f(r,m)$, of chance
associations found by misaligning the catalogue and MSX SiO targets,
where we plot separately, starting from the bottom, the distributions
of sources with \ks\ $< 9$, $9<$ \ks\ $<11$, and \ks\ $ > 11$ mag,
respectively.}
\end{centering}
\end{figure}

The distribution of angular separations of the \ks\ band counterparts
identified is shown in Fig.\ \ref{fig:denismsx.ps}.  The mean and
median separations are 3.0\arcsec\ and 2.7\arcsec\ with a standard
deviation of 1.8\arcsec, respectively, which is consistent with the
expected scatter due to positional uncertainties for the MSX and DENIS
catalogues.

To find the distribution of random chance associations we again
searched the nearest neighbours after shifting the coordinates of the
SiO targets by 30\arcsec. The resulting distribution of separations
and magnitudes is also shown in Fig.\ \ref{fig:denismsx.ps}. The real
associations are on average much brighter and closer than the chance
associations.  To compute the expected number of incorrect
identifications we divided the ``chance'' distribution in magnitude
bins, $(m,m+\Delta m)$, and in each bin computed the normalised
cumulative chance distribution of the separations, $f(m,r)$, i.e., the
fraction of all chance associations with those magnitudes and within a
radius $r$.  We then computed the sum $\Sigma f(m,r)$ over all
``real'' associations, each being characterised by $m$ and $r$. Thus,
$f(m,r)$ gives the probability for an association to be spurious, and
the sum over all sources yields the total expected number of spurious
identifications. When we consider only the brightest identifications
for the 154 sources with a \ks\ $< 9$ mag counterpart, only two
spurious identifications are expected. For the 34 possible
identifications with \ks\ $> 9$ mag, we would expect three to be
spurious. In total, we expect that about five of our identifications
may be incorrect.

Since for the fainter possible counterparts the chance of a false
identification is higher, we looked for brighter possible
identifications somewhat further away, which due to the lower surface
density of brighter sources might have a higher probability to be the
actual counterpart.  In a few cases (\#231, \#243, \#367, \#405, \#406,
and \#424) we found brighter sources somewhat further away but with a
lower value of $f(m,r)$ than the closest identification. This
suggested that these brighter sources were more likely to be the
correct counterparts, which we therefore retained.  For these MSX
targets with dubious near-infrared association, we additionally
examined other MSX sources in their surrounding and checked for
possible astrometric shifts between the DENIS and MSX coordinates which
could uniquely identify the correct near-infrared counterpart.
However, because of the low MSX source density ($\sim1$ source per
4\arcmin $\times$ 4\arcmin), only few associations could be confirmed
in this way.  Seven sources could only be associated with a \ks\ $>
11$ mag counterpart (\#162,\#273,\#363,\#403,\#417,\#423,\#443) within
15\arcsec\ (beam size of the IRAM telescope).


\subsection{2MASS data}
The Two Micron All Sky Survey (2MASS) has surveyed the entire sky at
near-infrared wavelengths from the Whipple Observatory on Mt. Hopkins,
AZ, and the Cerro Tololo InterAmerican Observatory (CTIO).  The
cameras observed simultaneously in the $J$ (1.25 $\mu$m), $H$ (1.65
$\mu$m) and \ks\ (2.17 $\mu$m) bands with a 10 $\sigma$ sensitivity
limit in uncrowded fields of 15.8, 15.1 and 14.3 mag, respectively
\citep[][]{beichman98,2massES}.  The absolute astrometry has typical
uncertainty of 0.1\arcsec\ (rms) and is based on the Tycho-2 and
UCACr10 catalogue (a new version of the USNO's ACT catalogue).  The
on-line 2MASS PSC (all sky data release) is accessible at the Infrared
Processing and Analysis Center (IPAC).  2MASS data was available for
all our positions.  We retrieved 2MASS sources for our 441 SiO target
positions and we searched for the closest positional match.  
Since the astrometric accuracy is better than 0.5\arcsec\ for the
ISOGAL targets but only $\sim$2\arcsec\ for the MSX targets, a
different distribution of angular separations is expected for the
MSX-2MASS and ISOGAL-2MASS associations.

\begin{figure}[h]
\resizebox{\hsize}{!}{\includegraphics{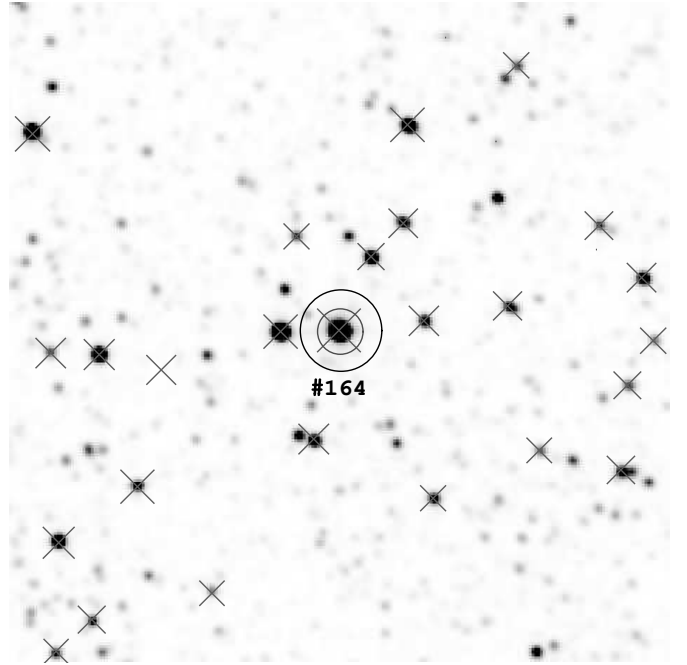}}
\caption{\label{fig:mm3.ps} 2MASS \ks\ image 230\arcsec $\times
230$\arcsec\ centred at the position of \#164. The black circle delimits
the main beam (FWHM) of the IRAM telescope used for the SiO maser
search.  A small gray circle marks the 2MASS association and finally
crosses mark sources detected by ISOGAL. In agreement with the
discussion in the text, all the ISOGAL and 2MASS positions are in
excellent agreement.  Note that the targeted source is the only
mid-infrared source falling inside the IRAM telescope beam.  }
\end{figure}

To avoid misidentification, because of the high source density of
the 2MASS survey and because the DENIS counterparts are mostly
brighter than 11 mag in \ks, we limited the 2MASS \ks\ magnitude to
\ks\ $< 11$.  However, seven sources in our sample, \#162, \#273,
\#363, \#403, \#413, \#421 and \#423, could be associated only with
2MASS sources fainter than \ks\ $= 11$ (those sources are also faint
in DENIS) within 15\arcsec, which is the beam size of the IRAM
telescope.  Positional associations were confirmed via overplotting
the 2MASS counterpart image with both the SiO targets and the DENIS
sources.  Finding charts were obtained for all the stars with 2MASS
images, an example of which is given in Fig.\ \ref{fig:mm3.ps}.  We
found 439 2MASS counterparts and missed only two. In fact, after image
inspections, the potential 2MASS counterparts for two sources, \#224
and \#298, were eliminated as their positions on the 2MASS images were
marked by artifacts.

\subsubsection{ISOGAL sample and 2MASS identifications}
\begin{figure}[h]
\resizebox{\hsize}{!}{\includegraphics{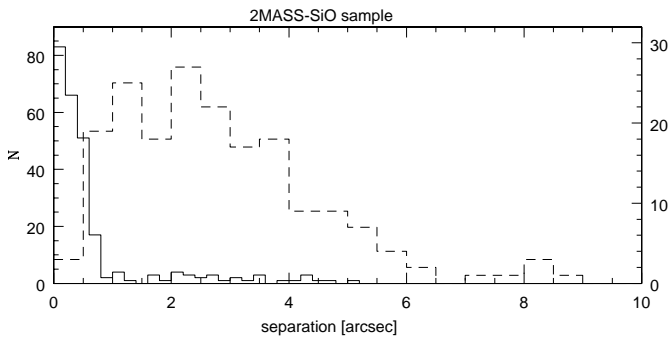}}
\caption{\label{fig:2massdist.ps} Distribution of the angular
separations between the SiO targets \citepalias[positions as
in][]{messineo02} and the 2MASS associations.  The continuum line
shows the distribution of the ISOGAL SiO targets and the corresponding
y-axis is on the left side.  The dashed line shows the distribution of
the MSX SiO targets and the relative y-axis is on the right side. }
\end{figure}

The mean, median separations between the ISOGAL SiO targets
\citepalias[positions as in][without rounding RA to one tenth of
second]{messineo02} and the 2MASS associations are 0.7\arcsec\ and
0.3\arcsec\ with a standard deviation of 1.0\arcsec, (see Fig.\
\ref{fig:2massdist.ps}).

There is a non-Gaussian tail at large separations.  We have
individually checked all the sources with separation larger than
3\arcsec\ and note that they all have \ks\ $< 6 $. We attribute the
large separation to saturation of the DENIS detector.  Saturated
pixels are an obstacle to the correct determination of the source
centroid and this affects the astrometry of saturated
stars. Furthermore, most of these bright sources do not have any $I$
associations and therefore the $J$/\ks\ astrometry is kept.  The mean,
median and standard deviation of the separations between the ISOGAL
SiO targets  and the 2MASS
associations with \ks$>6.5$ mag are 0.4\arcsec, 0.3\arcsec\ and
0.4\arcsec; while for 2MASS associations with \ks$<6.5$ mag they are
1.9\arcsec, 2.1\arcsec\ and 1.5\arcsec.

\subsubsection{MSX sample and 2MASS identifications}
We found 187 2MASS counterparts of MSX SiO targets.  The mean, median
and standard deviations of the separations between the MSX SiO targets
(using the MSX coordinates) and the 2MASS associations are 2.9\arcsec,
2.6\arcsec\ and 1.8\arcsec,
the same as the separations between the MSX SiO targets and the DENIS
associations (see Sect.\ \ref{msxden}).  This is again consistent with
the expected scatter due to positional uncertainties of sources in the
MSX and 2MASS catalogues.  The mean, median and standard deviation of
the separations between the MSX SiO targets (using the DENIS
coordinates) and the 2MASS associations are 0.6\arcsec, 0.4\arcsec\
and 0.7\arcsec, respectively.  The mean, median and standard deviation
of the separations in right ascension RA are $0.0$\arcsec,
$0.0$\arcsec\ and 0.5\arcsec; while the mean, median and standard
deviation of the separations in declination DEC are $-0.2$\arcsec,
$-0.1$\arcsec\ and 0.6\arcsec.

\subsection{SiO targets and  IRAS identifications}
\label{IRAS}
We checked for counterparts to our SiO targets in the Infrared
Astronomical Satellite Point Source Catalogue (IRAS PSC).  Toward the
galactic plane the IRAS survey is strongly limited by confusion at the
low resolution of the IRAS instruments (0.5\arcmin\ at 12$\mu$m),
especially in the longer wavelength bands.  Since much of the past
work on maser surveys of late-type stars is based on the IRAS data, it
may be useful to have an according comparison.
\begin{figure}[!]
\resizebox{\hsize}{!}{\includegraphics{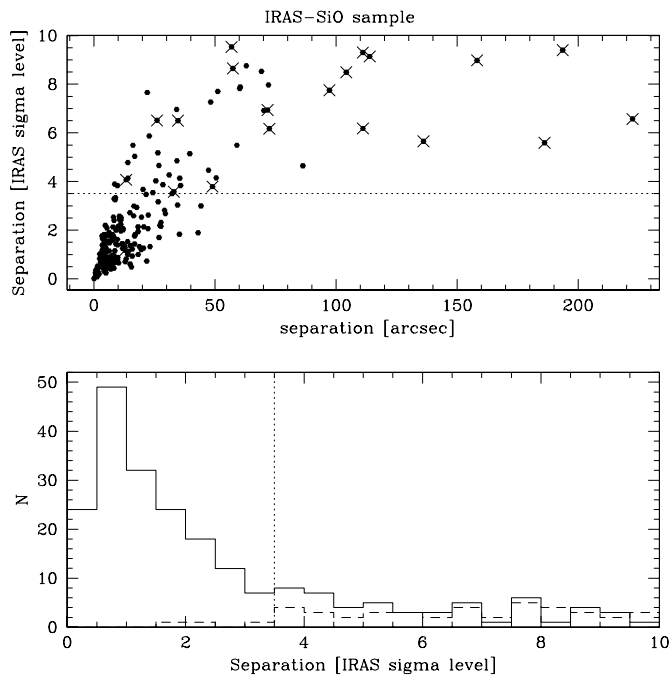}}
\caption{\label{fig:irasdist.ps} {\bf Upper panel:} Separation of the
 possible IRAS counterparts, expressed in IRAS sigma units, against
 angular separation. Crosses indicate IRAS sources with possible MSX
 counterparts closer than those associated with the SiO targets; these
 are likely to be unrelated to the SiO targets. The dotted horizontal
 line is the upper limit that is selected. {\bf Lower panel:} The
 continuum line shows the distribution of the separations (in sigma
 units) of the possible IRAS counterparts. The dashed line shows the
 distribution of the chance associations obtained shifting the source
 coordinates by 250\arcsec. The dotted vertical line is our chosen
 upper limit of 3.5 $\sigma$.}
\end{figure}

\begin{figure}[h]
\resizebox{\hsize}{!}{\includegraphics{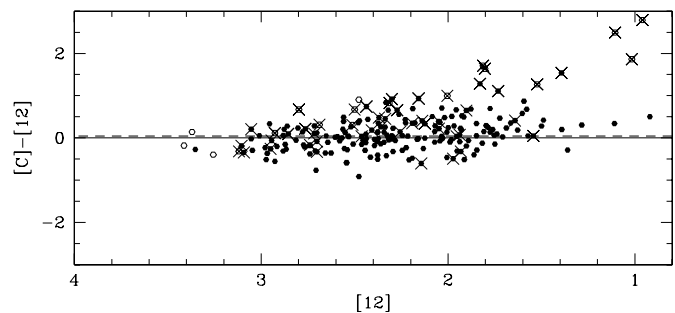}}
\caption{\label{fig:irascc.ps} Difference between the MSX magnitude in
the $C$ band, and the IRAS 12$\mu$m magnitudes, [12], (filled circles)
versus the [12] . The 18 open circles indicate sources without $C$
measurement, for which the 15$\mu$m ISOGAL measurement is plotted.
Crosses are likely spurious MSX-IRAS associations (angular separation
larger than 3.5$\sigma$).  The plotted IRAS magnitudes are $-2.5 \log
({F[{\mathrm {Jy}}]/28.3})$.  The continuous line is $C-[12]=0.00$ mag
and the dashed line is $C-[12]=0.05$ mag, which is the mean $C-[12]$
colour.  }
\end{figure}

We selected only the 165 IRAS associations within 3.5$\sigma$ error
ellipse of the IRAS PSC, to reduce the chances of spurious
associations to $\sim2$\% (3 sources).  A comparison of 12$\mu$m
fluxes of the prospective counterparts with the ISOGAL and MSX fluxes
(Fig.\ \ref{fig:irascc.ps}) shows a good agreement, confirming that
the IRAS identifications are proper.  About 35\% of our SiO targets
have counterparts in the IRAS PSC: 65\% of the MSX sample and 17\% of
the ISOGAL sample.  Of those, 96\% are detected at 12$\mu$m, 87\% at
25$\mu$m, 6\% at 60$\mu$m and 4\% at 100$\mu$m, and 56\% are reported
in the IRAS catalogue as variables (flag$> 80$).

\subsection{SiO targets and  visual identifications}

Within a radius of 5\arcsec\ to our SiO targets we searched for
possible visual counterparts in the Tycho 2 and USNO-A2.0 catalogues.
We found possible associations for 85 SiO targets, of which 27 are
validated by corresponding $I$ band counterparts.  Their R magnitudes
range from 10.9 to 17.9 and B magnitudes from 13.4 to 20.6.
Considering that the pulsation amplitude of Mira stars increases at
shorter wavelenghts and can be up to 8 mag in the visual
\citep{smak64}, all those visual stars are possible counterparts of
our SiO targets.  Most of them are located at latitude
$|b|<0.8^\circ$, but the extinction value inferred by their colours
are much smaller than the median of their surrounding stars
\citepalias{messineo03_3}. Therefore they are likely to be foreground
stars.  Two SiO targets, \#7 with
$(l,b,vel)=(-1.50^\circ,0.95^\circ,18.9$ \kms) and \#139 with
$(l,b,vel)=(0.31^\circ,-2.18^\circ,205.7$ \kms), have extinction value
inferred by their colours \Av$= 5.5$ and 3 mag, respectively,
consistent with the median extinction of their surrounding stars
\citepalias{messineo03_3}. Furthermore, the velocity of \#139 is
inconsistent with being a foreground star. Therefore we conclude that
they are likely located in the Galactic bulge, in regions of low
interstellar extinction.  In fact, \#139 is located in the optical
window $W0.2-2.1$ at $(l,b)=(0.2^\circ,-2.15^\circ)$
\citep{dutra02,stanek98}, and \#7 has an extinction value typical of
bulge ISOGAL fields with $b \sim+1$\degr\ \citep{ojha03}.

\subsection{The table}
Table \ref{table:rawdata} lists the infrared photometry of the SiO
targets.  The columns of Table \ref{table:rawdata} are as follows: an
identification number ($ID$), the same as in Table 2 and 3 of
\citetalias{messineo02}, followed by the Right Ascension ($RA$), and
Declination ($DEC$), in J2000 of the 2MASS counterpart; the DENIS
$(I,J,K_S)$ magnitudes, the ISOGAL $([7],[15])$ magnitudes, the 2MASS
$(J,H,K_S)$ magnitudes; the angular separation ($dis$), between the
adopted near-infrared position and the MSX position, the MSX
$(A,C,D,E)$ magnitudes, the IRAS 12 and  25 $\mu$m magnitudes and
finally a variability flag ($var$), defined as described in Sect.\
\ref{variability}.  An additional column is used for comments on
individual stars.

~~ \\ 
\medskip
\begin{sidewaystable*} 
\caption{\label{table:rawdata} Infrared counterparts of the SiO
targets.$^{\mathrm{*}}$ The identification numbers are the same as in
Table 2 and Table 3 of \citetalias{messineo02}.}  
{\footnotesize
\begin{tabular}{@{\extracolsep{-.04in}}rrr|rrrrr|rrr|rrrrr|rr|c|l}
\hline 
\hline 
    & &   &\multicolumn{5}{c}{\rm DENIS$$-$$ISOGAL}   &\multicolumn{3}{c}{\rm 2MASS} &   \multicolumn{5}{c}{\rm MSX}   &  \multicolumn{2}{c}{\rm IRAS}& &\\ 
\hline 
 {\rm ID}   & {\rm RA(J2000)} & {\rm DEC(J2000)}  &{\rm I} & {\rm J} & {\rm K} &
 [7] & [15]   &{\rm J} & {\rm H} & {\rm K} & {\rm dis} & {\rm A} & {\rm C} &
 {\rm D} &  {\rm E} & [12] & [25]&{\rm var}& Comments\\ 
\hline 
 &{\rm [hh mm ss]}   & {\rm [deg mm ss]}    &{\rm [mag]}   &	{\rm [mag]}    & {\rm [mag]}     & {\rm [mag]} &{\rm [mag]}  & {\rm [mag]}  & {\rm [mag]}  & {\rm [mag]} &{\rm \arcsec} &{\rm [mag]}    & {\rm [mag]}&{\rm [mag]}&{\rm [mag]}&{\rm [mag]}&{\rm [mag]}&& \\ 
\hline 
   1 & 17 31 40.98 & $$-$$32 03 55.9 &       & 12.77 &  8.28 &  5.16 &  3.48 & 12.53 &  9.66 &  8.10 &  3.6 & 5.40 &       &       &       &       &       &  1 &                         \\	
   2 & 17 36 42.18 & $$-$$30 59 11.7 &       & 14.92 &  8.16 &  4.89 &  3.06 & 14.67 & 10.42 &  8.09 &  1.0 & 3.82 &  2.60 &  2.35 &  1.17 &  2.97 &  1.58 &  0 &                         \\	
   3 & 17 37 07.29 & $$-$$31 21 31.3 & 18.29 & 12.22 &  7.53 &  4.99 &  3.55 & 12.42 &  9.30 &  7.60 &  2.5 & 4.74 &       &       &       &       &       &  2 &                         \\	
   4 & 17 37 29.35 & $$-$$31 17 16.6 &       & 15.01 &  7.97 &  4.31 &  3.44 & 15.41 & 10.59 &  8.08 &  3.6 & 4.94 &  3.25 &  2.93 &       &       &       &  2 &                         \\	
   5 & 17 38 11.78 & $$-$$31 46 27.0 &       & 11.78 &  7.02 &       &       & 11.62 &  8.70 &  7.02 &  4.5 & 3.36 &  2.36 &  2.19 &  1.25 &  2.51 &  1.56 &  2 &                         \\	
   6 & 17 38 12.49 & $$-$$29 39 38.5 & 17.36 & 11.77 &  8.06 &  5.26 &  2.89 & 11.73 &  9.69 &  8.29 &       &      &       &       &       &       &       &  1 &                         \\	
   7 & 17 38 17.07 & $$-$$29 42 32.4 & 12.65 &       &  6.42 &  4.63 &  3.02 &  7.75 &  6.22 &  5.38 &  1.8 & 3.92 &  3.18 &  2.92 &       &  2.44 &  1.57 &  2 &  M6, possible visual counterpart \\	
   8 & 17 38 29.01 & $$-$$31 26 17.5 &       & 10.68 &  6.85 &  4.04 &  2.93 & 11.01 &  7.94 &  6.29 &  3.5 & 4.35 &  3.04 &  2.86 &       &  2.80 &  0.95 &  2 &                         \\	
   9 & 17 38 32.50 & $$-$$31 20 42.7 &       & 13.94 &  7.67 &  4.74 &  3.44 & 14.14 &  9.99 &  7.75 &       &      &       &       &       &  1.80 &  0.55 &  2 &                         \\	
  10 & 17 38 35.69 & $$-$$29 36 37.2 & 19.15 & 11.21 &  7.33 &  4.59 &       & 12.58 &  9.80 &  8.12 &  1.5 & 3.76 &  2.76 &  2.57 &  1.34 &  2.81 &  1.30 &  2 &                         \\	
  11 & 17 39 37.28 & $$-$$30 08 51.6 & 16.94 &  9.99 &  6.63 &  4.57 &  3.09 & 10.02 &  7.90 &  6.67 &  3.3 & 4.27 &  3.14 &  3.02 &       &  2.86 &  0.69 &  2 &                         \\	
  12 & 17 40 57.23 & $$-$$29 45 31.4 &       & 11.60 &  7.16 &  4.07 &  2.70 & 11.87 &  8.87 &  7.48 &  2.4 & 4.32 &       &  2.58 &       &       &       &  2 &                         \\	
  13 & 17 41 16.81 & $$-$$31 38 10.6 &       & 13.54 &  8.30 &  4.98 &  3.05 & 14.04 & 10.49 &  8.43 &  2.2 & 3.98 &  2.75 &  2.42 &       &  2.98 &  1.33 &  1 &                         \\	
  14 & 17 41 26.93 & $$-$$29 30 47.0 &       & 11.80 &  7.39 &  4.48 &  3.56 & 12.00 &  9.17 &  7.39 &  1.2 & 5.13 &       &       &       &       &       &  2 &                         \\	
  15 & 17 41 31.31 & $$-$$30 00 18.9 & 16.43 &  8.61 &  6.36 &  3.24 &  2.03 &  7.73 &  5.73 &  4.64 &  4.3 & 3.09 &  1.99 &  1.97 &  1.04 &  1.95 &  0.23 &  1 &                         \\	
  16 & 17 41 36.86 & $$-$$29 29 31.0 &       & 12.23 &  7.06 &  3.79 &  2.00 & 12.69 &  9.53 &  7.58 &  2.8 & 3.43 &  2.43 &  2.14 &  1.02 &       &       &  2 &                         \\	
  17 & 17 41 37.41 & $$-$$29 32 05.7 &       & 12.26 &  7.37 &  4.60 &  2.85 & 12.18 &  9.08 &  7.30 &  0.9 & 4.52 &       &  2.61 &       &       &       &  2 &   no I band association \\	
  18 & 17 42 04.36 & $$-$$29 58 46.4 &       & 12.16 &  7.33 &  4.83 &  3.40 & 13.12 &  9.88 &  8.06 &       &      &       &       &       &       &       &  2 &                         \\	
  19 & 17 42 06.86 & $$-$$28 18 32.4 & 15.86 &  9.59 &  6.87 &  4.82 &  3.12 &  9.64 &  7.89 &  6.88 &  0.8 & 4.66 &  3.57 &  3.56 &  3.20 &       &       &  2 &                         \\	
  20 & 17 42 23.28 & $$-$$29 39 35.6 &       & 12.52 &  7.60 &  5.11 &  2.95 & 12.13 &  9.13 &  7.46 &  2.3 & 4.49 &  3.12 &  2.99 &       &       &       &  2 &                         \\	
  21 & 17 42 32.91 & $$-$$29 41 25.1 & 17.23 & 11.78 &  7.79 &       & 4.64 & 11.62 &  9.15 &  7.76 &       &      &       &       &       &           &       &  2 &                        \\	
  22 & 17 42 32.48 & $$-$$29 41 10.7 &       & 11.94 &  7.19 &  4.33 &  3.08 & 12.55 &  9.46 &  7.67 &  2.2 & 4.57 &  2.93 &  2.95 &       &       &       &  2 &                         \\	
  23 & 17 42 44.87 & $$-$$30 04 08.1 & 16.27 &  8.65 &  6.40 &  4.29 &  2.71 &  9.10 &  6.87 &  5.67 &       &      &       &       &       &       &       &  2 &                         \\	
  24 & 17 43 09.81 & $$-$$29 24 03.3 &       & 14.50 &  7.53 &  3.65 &  1.93 & 15.54 & 10.79 &  8.19 &  3.2 & 3.43 &  2.20 &  1.83 &  0.90 &       &       &  2 &                         \\	
  25 & 17 43 23.46 & $$-$$28 53 50.3 & 15.73 &  8.01 &  5.99 &  2.87 &  2.16 &  8.44 &  6.21 &  4.97 &  2.5 & 3.14 &  2.42 &  2.18 &       &       &       &  1 &                         \\	
  26 & 17 43 25.26 & $$-$$29 45 28.6 &       & 15.23 &  8.99 &  4.92 &  2.76 & 15.88 & 11.69 &  9.09 &  2.5 & 4.29 &  2.75 &  2.53 &       &       &       &  2 &                         \\	
  27 & 17 43 32.72 & $$-$$29 15 39.4 &       & 12.06 &  6.91 &  3.94 &  2.45 & 13.34 &  9.63 &  7.47 &  3.6 & 3.67 &  2.72 &  2.48 &  1.31 &       &       &  2 &                         \\	
 28 & 17 43 33.13 & $$-$$29 51 33.1 &       & 15.00 &  8.41 &  4.68 &  2.88 & 15.02 & 10.83 &  8.42 &       &      &       &       &       &  1.02 &  0.10 &  1 &                         \\	
  29 & 17 43 34.79 & $$-$$29 40 30.4 &       &       &  9.15 &  4.65 &  2.96 & 16.90 & 13.15 &  9.76 &       &      &       &       &       &       &       &  1 &                         \\	
  30 & 17 43 35.12 & $$-$$29 24 47.2 &       &       &  8.82 &  4.97 &  2.95 & 18.47 & 12.22 &  9.09 &  4.0 & 4.79 &  2.87 &  2.85 &       &       &       &  1 &                         \\	
\hline
\end{tabular}
}
\begin{list}{}{}
\item[$^{\mathrm{*}}$] The full table is available in electronic form
at the CDS via anonymous ftp to cdsarc.u-strasbg.fr (130.79.128.5) or
via \\ ${\rm http}://{\rm cdsweb.u-strasbg.fr/cgi}-{\rm
bin/qcat?J/A+A/(vol)/(page)}$.
\end{list}
\end{sidewaystable*}

\section{A comparison between the ISOGAL and MSX  samples}
\label{first}

\begin{figure*}[th!]
\begin{centering}
\resizebox{0.9\hsize}{!}{\includegraphics{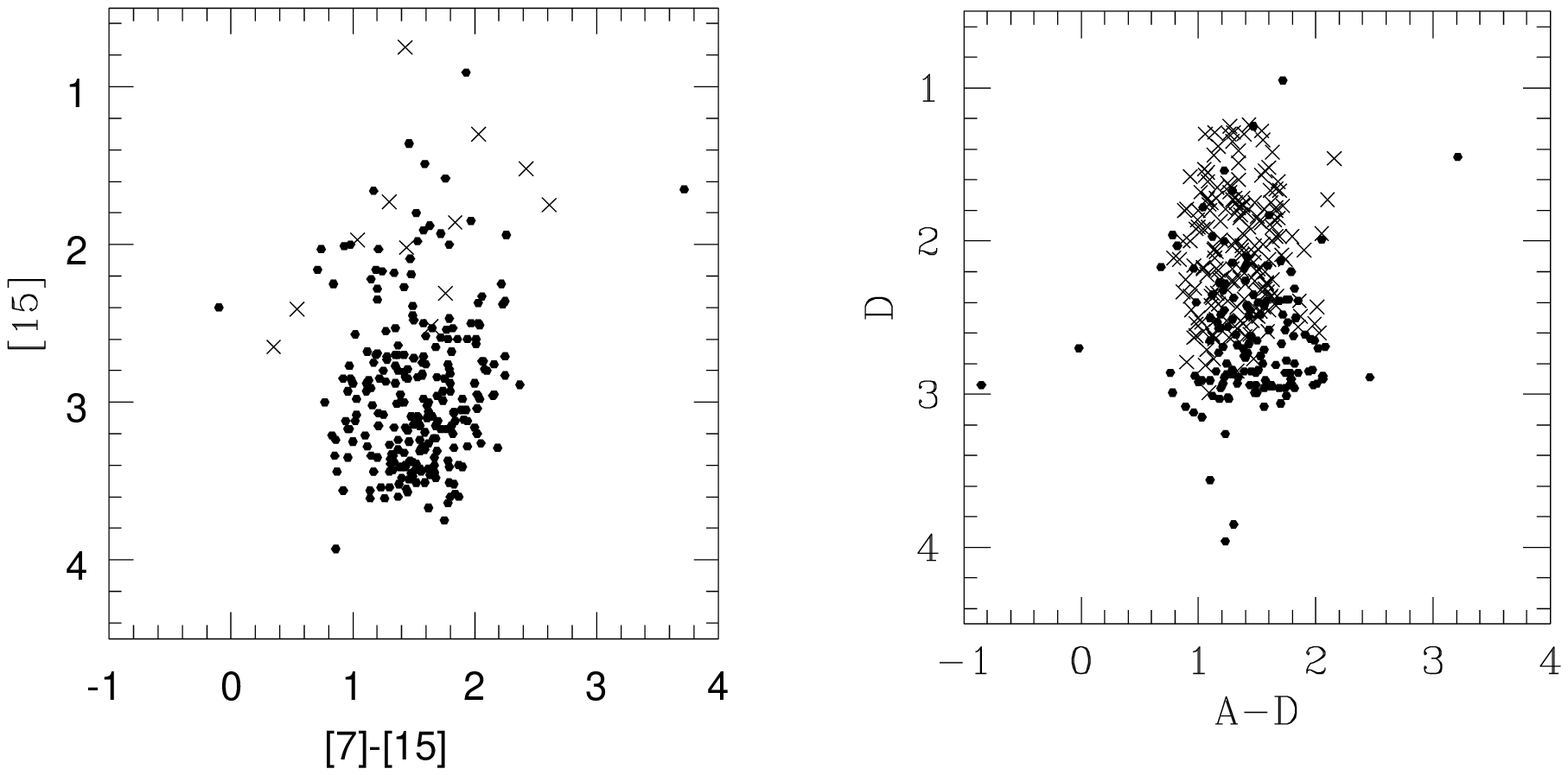}}
\caption{\label{fig:mir1.ps} Mid-infrared colour-magnitude
diagrams. {\bf Left-hand panel:} all data from ISOGAL.  {\bf
Right-hand panel:} all data from MSX.  Filled circles indicate targets
from the ISOGAL sample and crosses indicate targets from the MSX
sample. The two samples have similar mid-infrared colours, due to
selection criteria.  The numbers of MSX and ISOGAL targets between the
two panels vary as explained in Sect.\ \ref{msxiso-x}}
\end{centering}
\end{figure*}

A large variety of names exists to indicate oxygen-rich AGB stars
characterized by different pulsation properties and/or mass-loss rate:
semi-regular (SR) stars and Mira stars (H$\alpha$ in emission, visual
pulsation amplitude larger than 2.5 mag), large amplitude variables
(LAV), long period variable (LPV) stars (when their periods are longer
than 100 days), and OH/IR stars (with 1612MHz OH maser emission).  In
the IRAS colour-colour diagram the oxygen-rich AGB stars are
distributed on a well-defined sequence of increasing shell opacity and
stellar mass-loss rate (e.g. Habing 1996), which goes from SRs and
Miras with the bluest late-type colours and the 9.7 $\mu$m silicate
feature in emission, to the coldest OH/IR stars with the reddest
colours and the 9.7 $\mu$m silicate feature in absorption. The
sequence of increasing shell opacity corresponds also to an increasing
\ks$-[15]$ or \ks$-[12]$ colour
\citep[e.g.][PaperI]{ojha03,whitelock94,olivier01}.

SiO maser emission is generated in the envelopes of mass-losing AGB
stars, close to their stellar photospheres and it occurs more
frequently towards oxygen-rich Mira stars than towards other AGB stars
(including SR and OH/IR stars) \citep{bujarrabal94,nyman93}.
Therefore, for our 86 GHz SiO maser survey we selected the brightest
sources at 15 $\mu$m with colours of Mira-like stars
\citepalias{messineo02}.  The ISOGAL and MSX samples were selected to
have similar 7 and 15 $\mu$m colours, as shown in Fig.\
\ref{fig:mir1.ps}.  For the ISOGAL sample a range of intrinsic
(\ks$-[15])-$colour was selected (see Sect.\ \ref{isogal}), but for
the MSX sample no near-infrared counterparts were available at the
time of the observations, and thus no \ks\ magnitudes.  As an
alternative to the (\ks$-[15])$ criterion, for the MSX sample we
imposed an upper limit to the ratio of the fluxes in the $E$
(21$\mu$m) and $C$ (12$\mu$m) bands ($C-E < 1.55$ mag).  Both criteria
were defined in order to avoid non-variable objects and thick envelope
objects, but these criteria are not equivalent.  The emission in the
\ks\ band is dominated by the stellar emission attenuated by the
circumstellar absorption, while circumstellar dust emission
contributes strongly to the mid-infrared radiation.

AGB stars with thin envelopes have $C-E < \sim1.5$ mag, while sources
with $C-E> \sim1.5$ mag are AGB stars with thick envelopes, post-AGB
stars and young stars \citep{sevenster02I,lumsden02}.  Our sample of
SiO targets includes very few sources with $C-E> 1.5$.

\begin{figure}[h]
\resizebox{0.9\hsize}{!}{\includegraphics{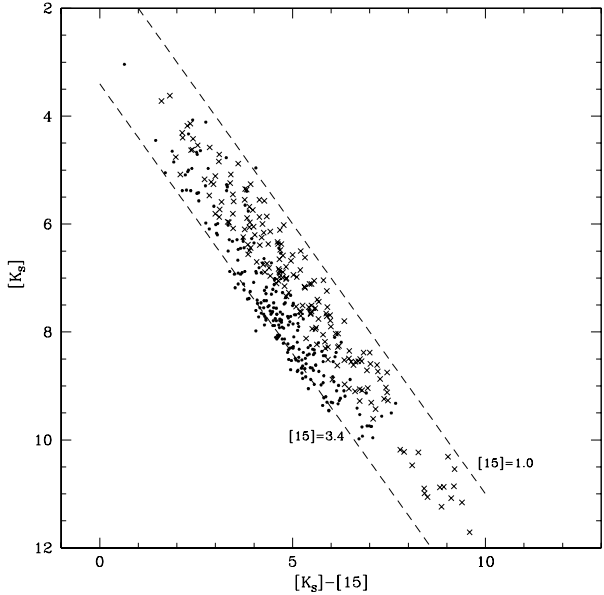}}
\caption{\label{fig:mir2.ps} 2MASS \ks\ versus (\ks$-[15])$ or
(\ks$-D)$. Filled circles indicate objects from the ISOGAL sample, for
which we plot (\ks$-[15])$; and crosses indicate objects from the MSX
sample, for which we plot (\ks$-D)$.  The two dashed lines correspond
to $[15]$ = 3.4 and 1.0 mag.}
\end{figure}

The \ks$-[15]$ and \ks$-[12]$ are good indicators of mass-loss rate
for AGB star with shells at few hundred Kelvin. All Miras have redder
\ks$-[12]$ than non-Mira stars from 1.8 mag up to 6 or even 14 mag
\citep[e.g.][]{olivier01,whitelock94}.  Thick-envelope OH/IR stars
have typically \ks$-[15] >4$ \citep{ortiz02}.  The ISOGAL SiO target
sample is characterised by a smaller range of (\ks$-[15])$ or
(\ks$-D)$ values than the MSX sample, as shown in Fig.\
\ref{fig:mir2.ps}, suggesting that the MSX sample includes a tail of
sources with optically thick envelopes.  However, this will be
verified after correction for extinction
\citepalias{messineo03_3,messineo03_4}.

Figure \ref{fig:mir2.ps} shows a strong correlation, between \ks\ and
(\ks$-[15])$ or (\ks$-D)$, which is due to the way we selected our
original sample.  In fact, due to the general correlation of SiO maser
emission and infrared luminosity (Bujarrabal et al.\ 1987), we
selected only sources with $[15] <$ 3.4 in order to be able to detect
the expected SiO line.  The $[15]$ and $D$ magnitudes range from
$\sim3.4$ to $\sim1.0$.  This narrow range generates the correlation
seen in Fig.\ \ref{fig:mir2.ps}.  

Figure 10 also shows that the two samples, ISOGAL and MSX, overlap
largely, although there are some minor systematic differences.  There
is a vertical shift between the ISOGAL and MSX sequence, with the MSX
sources brighter for a given colour.  This effect is due to the
different sensitivity between the MSX $D$ band and ISOGAL surveys.
MSX targets have on average a brighter $D$ (or [15]) magnitude than
the ISOGAL targets (see Fig.\ \ref{fig:mir1.ps}).  This fact
translates in differences in distance between the two samples and
suggest that the MSX stars are closer on average; the two samples are
also distributed differently in longitude (see Fig.\ \ref{fig:lb.ps}).

On the basis of this comparison in the following we combine the
results obtained with the ISOGAL sample with those obtained with the
MSX sample, taking into account that there are some minor differences
between the two samples.

\section{Some other remarks on the SiO targets}
\label{remarks}
\subsection{SIMBAD search}
\label{simbad}
A SIMBAD search revealed some extra information about our SiO
targets.

Sources \#7, \#286 and \#303 are included in the catalogue of
late-type stars in the inner Galactic region by \citet{raharto84} as
spectral types M6, M7 and M6.5, respectively.

\#153 is a well known Mira star, TLE 53, with a period of 480 d,
located in  Baade's window \citep[e.g.][]{glass95}.
Our sample also includes 15 LPVs found by \citet{glass01} within
0.3\degr\ from the Galactic center and 19 candidate variable stars
from the list of \citet{schultheis00} \citepalias[listed in Table 2
and 3 of][]{messineo02}.

A few sources are given in the literature as possible red supergiants
or extremely luminous AGB stars: \#25, \#32, \#92 and \#295 correspond
to sources \#6, \#8, \#31 and \#5 of \citet{nagata93}, respectively;
\#356 is classified as bulge M supergiant by \citet{raharto91} and
\citet{stephenson92} also lists it among distant luminous early type
stars. 

\#127 (IRAS 17500$-$2512), \#178 (IRAS 18040$-$2028), \#188 (IRAS
18060$-$1857), \#252 (IRAS 18285$-$1033), \#265 (IRAS 18367$-$0507)
and \#434 (IRAS 18415$-$0355) are listed by \citep{kwok97} among
sources detected with the IRAS Low Resolution Spectrometer, in the
range 8-23 $\mu$m and with a resolution, $\lambda/\Delta \lambda,
\sim20-40$.  Four spectra are noisy or incomplete, while the spectra
of \#178 (IRAS 18040$-$2028), \#434 (IRAS 18415$-$0355) are classified
as featureless; they are probably evolved stars with negligible
amounts of circumstellar dust.  For those two objects, at longitudes
9.7 and 28.6\degr\ respectively, we also compute a moderate mass-loss
rate of $5-7 \times 10^{-7}$ M$_\odot$ yr$^{-1}$
\citepalias{messineo03_3}.
 
\#164, IRAS 17590$-$2412, is classified as a Li K giant star by
\citet{delareza97}.  There is a significant difference between the SiO
heliocentric velocity $V_{hel}=+4.4\pm1.0$ \kms\ and the optical
heliocentric velocity $V_{hel}=-14.5\pm1.0$ \kms\ \citep[][de La Reza
{\it priv. communication}]{torres99}.  Since the SiO maser velocity is
usually coincident with the stellar velocity within few \kms\
\citep[e.g.][]{habing96}, we suggest that the SiO emitter is not
associated with the G8II star, IRAS 17590$-$2412/PDS 482, which
however is the only mid-infrared source within the IRAM beam and the
association between the ISOGAL and DENIS source is of excellent
quality (flag=5).

\#189, IRAS 18059$-$2554, is given by \citet{lynch90} as a possible
member of the globular cluster NGC6553. The stellar line-of-sight
velocity, obtained through the SiO maser line, is 161.2 \kms.  Because
the cluster mean line-of-sight velocity is 7 \kms\ with $\sigma = 14$
\kms\ \citep{coelho01}, we conclude that IRAS 18059$-$2554 is not a
member of the cluster.

Twenty-eight of our 86 GHz SiO targets were previously observed for 43
GHz SiO maser emission.  These are discussed in Sect. 4.5 of
\citetalias{messineo02}.  For the strongest 86 GHz maser sources
within 2.2\degr\ of the Galactic Centre from \citetalias{messineo02}
we recently used the Very Large Array (VLA) to observe the two 43 GHz
SiO maser lines (v=0 and v=1) simultaneously \citep{sjouwerman03}.

We excluded from our SiO maser survey the OH/IR stars detected by
\citet{sevenster97a, sevenster97b, sevenster01}, \citet{sjouwerman98}
and \citet{lindqvist92}.  However, due to intrinsic source variability
and the limited sensitivity of the Sevenster et al. surveys, we still
included 4 OH/IR stars, as found with a SIMBAD search: \#181, \#226
and \#257, all with detected SiO maser emission, coincide in position
and velocity with OH9.84+0.01, OH17.43$-$0.08 and OH25.05+0.28,
respectively \citep{blommaert94}; \#409 (IRAS 18142$-$1600), not
detected in our SiO survey, corresponds to OH \#280 (OH14.805+0.150)
listed by \citet{lintel89}.


\subsection{Is the targeted star the actual SiO emitter?} \label{beam} Using
the IRAM-30m telescope, we detected 86 GHz SiO maser emission toward 268 of our
targeted positions, where three positions showed a double detection (\#21-\#22,
\#64-\#65 and \#78-\#79). \footnote{ Throughout the paper for statistic
computations only one infrared source, the targeted, is considered for these
three positions.  A second mid-infrared source is found within the IRAM beam at
the position of the SiO sources \#78 and \#79 (separation $\sim12$\arcsec), and
at the position of the SiO sources \#21 and \#22 (separation of
$\sim15$\arcsec) \citepalias[see][]{messineo02} and is given in Table
\ref{table:rawdata}. }  To confirm that our targets are the actual SiO
emitters, we need to examine all other possible objects falling inside the
29\arcsec\ (FWHM) IRAM beam.  Due to the correlation between intensity of the
maser line and mid-infrared brightness \citep{bujarrabal87}, we can restrict
the analysis to only mid-infrared objects.  Considering the 379 SiO targets
with an MSX identification, we found in all but one case only one target inside
the 86 GHz beam.  Considering the 267 SiO targets with an ISOGAL
identification, 89\% are unique ISOGAL objects inside the main beam.  SiO maser
emission was detected towards 19 of our 29 positions with more than one ISOGAL
source in the beam and in all cases the targeted star is the brightest 15
$\mu$m source.


\begin{figure}[h]
\resizebox{\hsize}{!}{\includegraphics{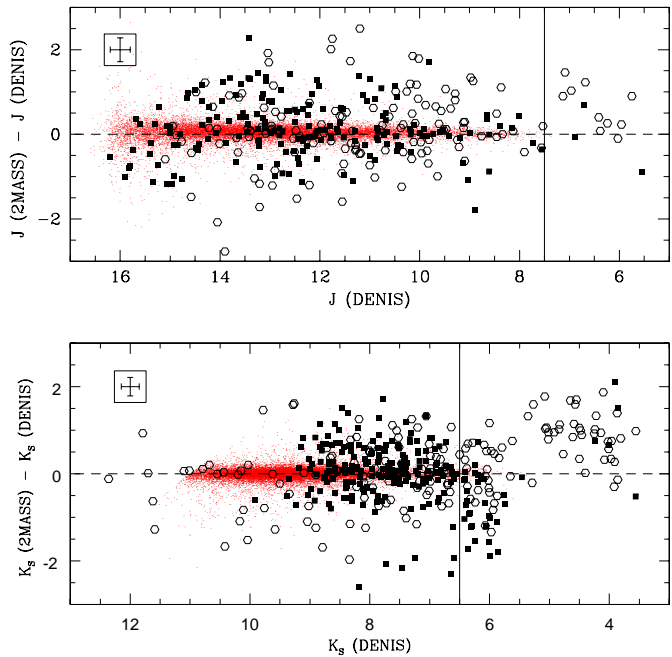}}
\caption{\label{fig:tot.ps} {\bf Upper Panel:} Difference between 
2MASS and DENIS $J$ magnitudes versus the DENIS $J$ magnitude.
Filled squares represent the ISOGAL SiO targets.  For comparison small dots
show ID-PSC and 2MASS associations obtained in several ISOGAL fields,
which have a distribution of their $J$ variations consistent with a
gaussian distribution centred at zero.  No correction for offset in
the photometric zeropoint was applied.  Open circles represent the MSX
targets, which have indications of variability at mid-infrared wavelengths.
The error bars shown within the box are $\pm1\sigma$ mag.  The
vertical line indicates the DENIS saturation limit.  {\bf Lower
Panel:} Comparison of the 2MASS and DENIS \ks\ magnitudes versus the
DENIS \ks\ magnitude. Symbols are as in the top panel.}
\end{figure}
\section{Variability}
\label{variability}
The 86 GHz SiO maser line intensity is observed to be stronger in
O-rich Mira stars than in other types of AGB stars
\citep{nyman93,bujarrabal94}.  To increase our chance of detecting the
SiO maser emission, we tried to select strongly variable sources
\citepalias[][and references therein]{messineo02}.  Therefore, our
selected MSX targets all have an indication of variability in the $A$
band.  Unfortunately, the ISOGAL database does not provide direct
information on variability. However, from their position in the
ISOGAL/DENIS (\ks$_0-[15]_0$ vs. $[15]_0)$ diagram we were expecting
that our ISOGAL SiO targets were mainly large amplitude variables
\citepalias[Sect.\ 4.2 of ][]{messineo02}.  For 62 ISOGAL targets out
of the 191 which have an MSX counterpart, there is an indication of
variability in the $A$ band (or for 94 in at least one of the $A$,
$C$, $D$, and $E$ bands).

\subsection{Variability information from DENIS and 2MASS data}
The comparison between the near-infrared photometry obtained during
the course of the DENIS survey and that of 2MASS again confirms our
hypothesis for variability in the ISOGAL sample.  

The $J$ and \ks\ filters used by DENIS and 2MASS are similar and
therefore the measurements obtained during the course of the DENIS and
2MASS surveys are directly comparable. For non variable sources, the
differences between the $J$ magnitude of DENIS and 2MASS and the \ks\
magnitude of DENIS and 2MASS are smaller than 0.15 mag \citep[][ and
present work]{delmotte02,schultheis01}.  

The DENIS observations were performed between 1996 and 2000, while the
relevant 2MASS observations were performed between 1998 and 2000; AGB
variables have periods from 50 to 1000 days, therefore the interval of
time between the observations makes it possible to derive variability
information.  

\begin{figure}[h]
\begin{centering}
\resizebox{0.7\hsize}{!}{\includegraphics{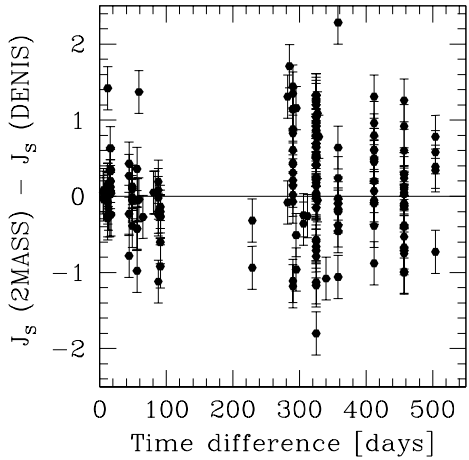}}
\caption{\label{fig:jul.ps} The difference of the 2MASS and DENIS $J$
magnitude versus the time between the 2MASS and DENIS observations.}
\end{centering}
\end{figure}

For each of the 61 ISOGAL fields containing our SiO targets we
retrieved the corresponding 2MASS sub-catalogue, and cross-correlated
the ID-PSC and the 2MASS point source positions.  For our ISOGAL
sample the difference between the 2MASS and DENIS $J$ and \ks\
magnitudes is shown as a function of the corresponding DENIS magnitude
in Fig.\ \ref{fig:tot.ps}.  Figure \ref{fig:jul.ps} shows the
difference between the 2MASS and DENIS $J$ magnitudes plotted against
the time between the 2MASS and DENIS observations.  The distribution
of the magnitude variations of the ISOGAL targets is different
compared to that of random field objects.  The Kolmogorov-Smirnov test
gives a zero probability for the ISOGAL targets and field stars to be
extracted from the same population.  For 55\% of our ISOGAL stars, the
difference between both the 2MASS and the DENIS $J$ and the 2MASS and
the DENIS \ks\ magnitudes is larger than 3 times the dispersion
measured in the corresponding field.  Therefore, our sample contains
mostly variable stars.

\begin{figure}[h]
\resizebox{\hsize}{!}{\includegraphics{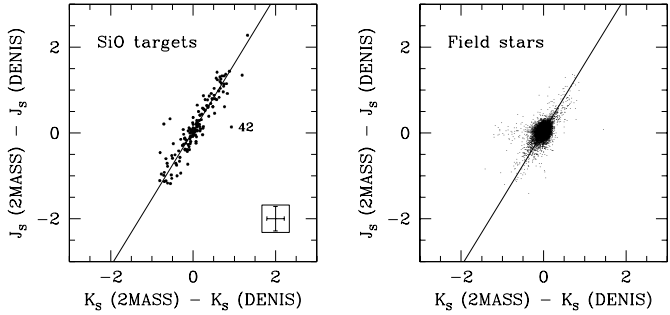}}
\caption{\label{fig:jk.ps} Difference between the 2MASS and DENIS $J$
magnitudes versus the difference of the 2MASS and DENIS \ks\
magnitudes. {\bf Left panel}: SiO maser targets.  Upper limit
measurements and measurements affected by saturation are excluded.
The photometry of \#42 is affected by blending with another source.
The continuous line shows our best fit.  Within the box the typical
error for sources in our range of magnitude is drawn.  {\bf Right
panel}: for comparison, field stars with $J<14$ mag.}
\end{figure}

Due to the simultaneity of the $J$ and \ks\ measurements in both the
DENIS and 2MASS surveys, a correlation is expected between the
variation in the $J$ magnitude ($\Delta J$) and in the \ks\ magnitude
($\Delta$\ks). As shown in Fig.\ \ref{fig:jk.ps} such a correlation
exists.  A linear least squares fit yields
$$\Delta J =1.57 (\pm 0.03) \times \Delta K_S + 0.05 \pm (0.01).$$ The
relative pulsation amplitude in the \ks\ band is $\sim60\%$ of the
relative amplitude in $J$ band.

\begin{figure}[h!]
\begin{centering}
\resizebox{0.5\hsize}{!}{\includegraphics{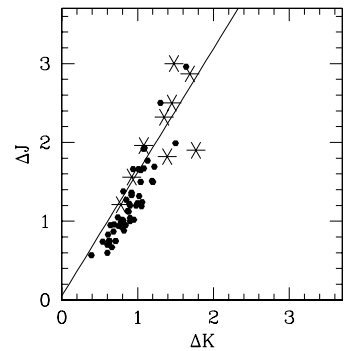}}
\caption{\label{fig:whitelock.ps} Pulsation amplitude in $J$ band
versus amplitude in $K$ band for two samples of oxygen-rich variable
stars: one in the solar neighbourhood from \citet{olivier01} (starred
points) and another in the South Galactic Cap from \citet{whitelock94}
(filled circles). Superimposed is our best fit from Fig.\
\ref{fig:jk.ps}.  }
\end{centering}
\end{figure}

For comparison, in Fig.\ \ref{fig:whitelock.ps}, we also show the
relation between the pulsation amplitudes in the $J$ and $K$ SAAO
bands for two different samples of oxygen-rich stars in the solar
neighbourhood \citep{olivier01} and the South Galactic Cap (SGC)
\citep{whitelock94}.  The two samples have a different period
distribution; most of the stars from the solar neighbourhood sample
have periods between 500 and 700 d, while most of the SGC stars have
periods between 150 and 450 d.  Overplotting our best-fit, we see that
it aligns well with the distribution of the two samples of LPVs. 

A monitoring program of the near-infrared magnitudes of our SiO maser
sample will provide pulsation periods and estimates of the source
distances through the period-luminosity relation.

\subsection{Variabity flag}
All the available information on variability, from MSX data
(variability in at least one of the MSX band), IRAS data (variability
flag $>50$), DENIS-2MASS data ($\Delta J > 3 \sigma(J)$ and $\Delta K
> 3 \sigma(K)$), or from \citet{glass01}, from \citet{schultheis01}
and from \citet{glass95}, is summarised defining a flag ($var$) which
is 2 when variability is detected in at least one of the datasets, 1
when variability is not detected but not all the datasets were
available and 0 when variability is not detected in any of the
datasets.  All the 188 MSX targets have the variability flag equal to
2 (due to selection).  Among the 253 ISOGAL targets 150 have the flag
equal to 2, 81 equal to 1 and 22 equal to 0.

There is no correlation between the variability indication and the
detection of SiO maser emission.

\section{The distribution of the SiO targets in
the IRAS two-colour diagram}\label{irascolours}
\begin{figure}[h]
\resizebox{\hsize}{!}{\includegraphics{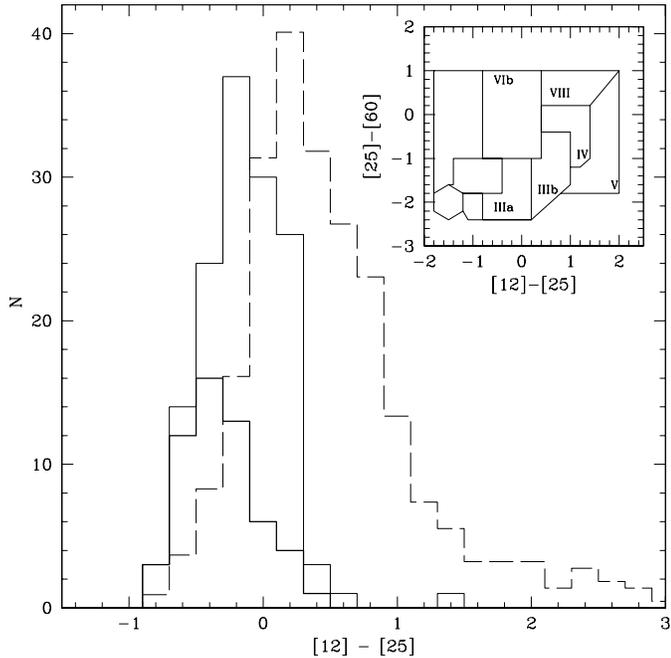}}
\caption{\label{fig:irash.ps} Distribution of the IRAS $[12]-[25]$
colours, $-2.5 \log(F_{12}/F_{25})$.  The continuous thin line
indicate the distribution of our SiO targets; the thick line indicates
the distribution of SiO targets with \ks\ $< 6.5$ mag.  For
comparison, the dashed thin line indicates the distribution of the
colours of OH/IR stars of \citet{sevenster02I}.  Upper limit flux
densities are excluded.  In the upper right corner, we recall the IRAS
two-colour diagram.}
\end{figure}

We cannot determine the distribution of our target sources in the
two-colour IRAS diagram of \citet{vanderveen88} since for most of our
sources we have only upper flux limits at 60 $\mu$m.  Figure
\ref{fig:irash.ps} shows the distribution of the IRAS $[12]-[25]$
colours and recalls the regions of the IRAS two-colour diagram that
separate different classes of evolved stars with circumstellar
envelopes, from the bluer Mira to the thick envelope OH/IR stars.  The
$[12]-[25]$ colours of our selected sources range from $-1.0$ to
$0.4$, peaking at $-0.2$, corresponding mostly to regions IIIa and
VIb.  Region IIIa represents sources with moderate dust emission,
being populated mostly by oxygen-rich stars with silicate emission
\citep{vanderveen88,kwok97}.  Region VIb, whith 60$\mu$m excess,
contains a mixed population of early type stars with line emission and
planetary nebula; however this region is scarcely polulated.
Considering the distribution of the IRAS good quality sources, only
5\% of them are located in region VIb.  We conclude that our targets
are mostly Mira stars with moderate mass-loss rate, in agreement with
the selection criteria. For comparison, the distribution of the
$[12]-[25]$ colours of the OH/IR stars of \citet{sevenster02I} is also
shown in Fig. \ref{fig:irash.ps}.  OH/IR stars are distributed over a
larger and redder colour range, although partially overlapping with
the colours of the SiO sample; they can have significantly ($\sim1$
mag) redder colours than the SiO targets. This can not be accounted
for by interstellar reddening, which is only 0.1-0.2 mag for \Av\
$\sim20-30$ mag.  The stars in our sample brighter than \ks\ $< 6.5$
mag are bluer than those with \ks\ $> 6.5$ mag. They are likely to be
mostly foreground stars with thinner shell, but still IRAS detectable
due to their proximity.

\#177 (IRAS 18039$-$2052) is the only source with $[12]-[25] > 1.0$,
which resembles those colours of an embedded young stellar object.
However, methanol and water maser emission was unsuccessfully searched
for toward this source
\citep{macleod98,palla91,molinari96}. Furthermore, the detection of
SiO maser emission confirms that the source is a late-type star, since
SiO maser emission is extremely rare in star-forming regions, with
only three (extremely luminous) sources detected to date
\citep[e.g.][]{engels89,ukita87,snyder74}. Though both the 12$\mu$m
and 25$\mu$m IRAS flux densities have good quality, their association
is unreliable. In fact, the corresponding MSX source has measurements
in the $A$, $C$ and $D$ bands consistent with the 12 $\mu$m IRAS
detection, but is not detected at 21 $\mu$m.

\section{MSX colour-colour diagrams}
\label{msxcolours}
\begin{figure}[!]
\resizebox{\hsize}{!}{\includegraphics{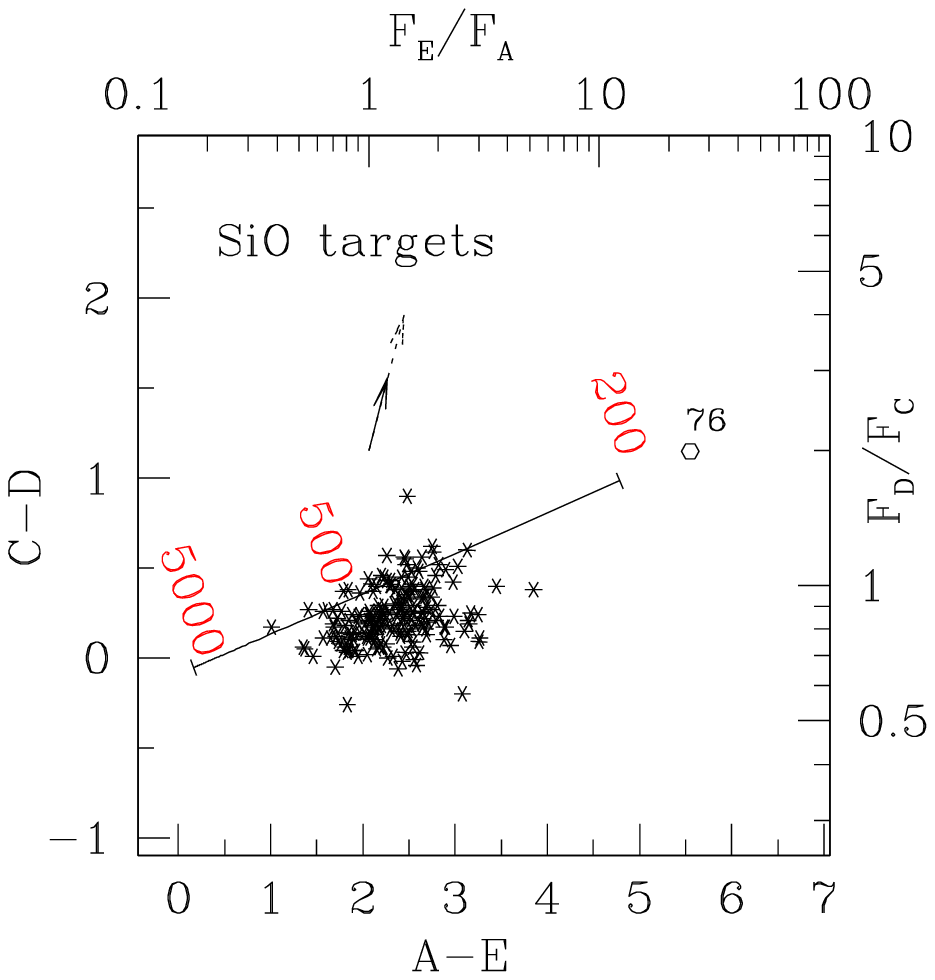}}
\caption{\label{fig:lum.ps} MSX colour-colour plot for our SiO maser
targets, similar to Fig.\ 5 of \citet{lumsden02}.  Stars represent AGB
stars, open circles early post-AGB stars, following the classification
of \citet{sevenster02I}.  Black body spectra follow the continuous
line.  A reddening vector for \Av\ $=40$ and extinction law from
\citet{Mathis90} is given by the solid arrow, while another one
obtained using the extinction law of \citet{lutz99} is represented by
the dashed arrow.  Note that the two vectors point in the same
direction. Our SiO targets have colours similar to the objects with
9.7 $\mu$m silicate emission in \citet{kwok97} and \citet{lumsden02}.
Object \#99 is outside of the plotted region ($A-E=3.01$ mag,
$F_E/F_A=2.4$; $C-D=-2.99$ mag, $F_D/F_C=0.04$) (see also text). }
\end{figure}

\begin{figure}[!]
\resizebox{\hsize}{!}{\includegraphics{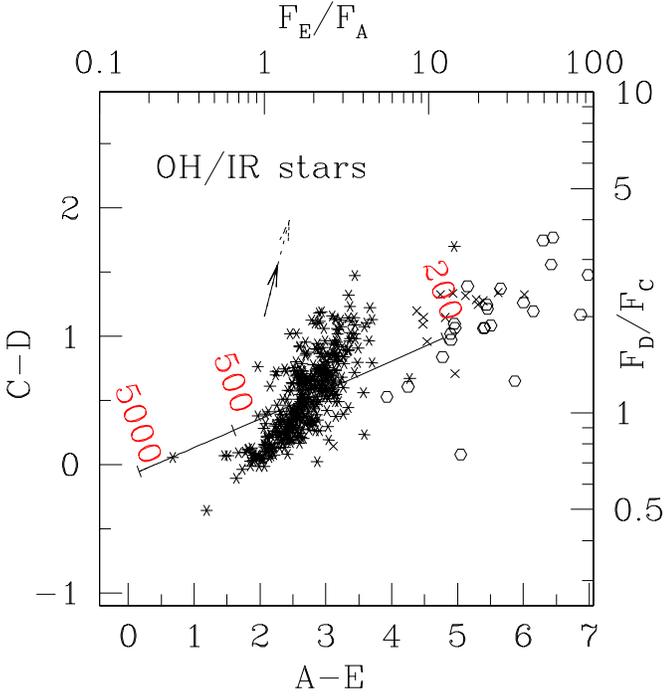}}
\caption{\label{fig:lumoh.ps} MSX colour-colour plot as in Fig.\ 5 of
\citet{lumsden02} for the OH/IR sample of \citet{sevenster02I}.
Symbols are as in Fig.\ \ref{fig:lum.ps}, plus crosses which
represent late post-AGB stars, following the classification of
\citet{sevenster02I}}
\end{figure}

\citet{sevenster02I} analysed the mid-infrared properties of her OH/IR
sample using IRAS and MSX data.  Studying a possible correspondence
between regions in the two-colour IRAS diagram and the MSX $A-C$
versus $D-E$ plane she suggested that the MSX diagram can
distinguish between the AGB and the post-AGB phases.  The transition
from a blue ($< 1.8$) to red ($> 1.8$) $A-C$ colour may correspond
to a transition off the AGB to proto-planetary nebulae: the star had
its last thermal pulse, and ceased to be variable.  The transition
from a blue ($< 1.5$) to a red ($> 1.5$) $D-E$ colour indicates a
later evolutionary transition, when mass-loss starts to drop down of
several order of magnitudes and there is the onset of the fast wind.
Most of our SiO targets with a clear MSX counterpart show $A-C<1.8$
and $D-E<1.5$, as expected for AGB stars, and comparable to the bulk
of Sevenster's OH/IR sample. 

According to Sevenster' criteria, only two objects, \#76 and \#99,
which are both SiO maser emitters, are likely to be post-AGB stars.
The odd colour of \#99 is due to an extremely high flux density of 39
Jy measured in the $C$ band, while the flux density in both the $A$
and $D$ bands is only $\sim1.7$ Jy.  However, the inspection of the
MSX $C$ band image does not confirm the presence of a such bright
object and we conclude that the $C$ photometry (despite its good flag)
of \#99 is unreliable. Source \#76, LPV 12-352 \citep{glass01}, is
located on a region of extended emission which is associated with a
star forming region \citep{schuller02}; this could have affected the
mid-infrared colours of \#76.  In conclusion: the anomalous position
of the sources \#99 and \#76 is probably due to the assignment of
incorrect magnitudes.

The $A-C$ versus $D-E$ colour diagram is useful to locate post-AGB
stars, which have redder colours due to their colder envelopes.
However, this diagram can hardly distinguish between different
thickness of the envelopes of AGB stars.  Using a sample of IRAS
sources with IRAS low resolution spectra \citep{kwok97},
\citet{lumsden02} showed that a different combination of MSX filters
can distinguish between circumstellar envelopes with silicate feature
at 9.7 $\mu$m in emission and in absorption.  Miras and OH/IR stars
with the silicate feature in emission are located below the black body
line in the $C-D$ versus $A-E$ diagram, while OH/IR stars with
silicate feature in absorption lie above this line.  In optically
thick envelopes, self-absorption causes a decrease of the flux in the
$C$ and $A$ bands, leading to an increase in $C-D$.  Figure
\ref{fig:lum.ps} shows that our SiO targets are distributed like
objects with the silicate feature in emission. We note that the MSX
two colour plot is not corrected for reddening.  However, the
reddening correction makes the sources bluer in the $C-D$ colour
independently of the adopted extinction law.  Furthermore, in the
$C-D$ versus $A-E$ diagram the distribution of the ISOGAL and MSX
samples are similar.  For comparison, in Fig.\ \ref{fig:lumoh.ps} the
MSX colour-colour plots of the Sevenster's OH/IR stars is shown, which
are distributed over a wider and redder range of colours.

\section{Conclusion}\label{conclusion}
To increase the number of line-of-sight velocities measured toward the
inner Galaxy we searched for 86 GHz SiO maser emission in a sample of
441 late-type stars. While the radio survey was described elsewhere
\citep{messineo02}, this paper presents the infrared photometry of all
our SiO targets as derived from the various accessible infrared
catalogues:  DENIS, 2MASS, MSX, ISOGAL and IRAS catalogues.

As described in \citetalias{messineo02}, we initially selected the SiO
targets from the ISOGAL and MSX catalogues on the basis of their near-
and mid-infrared colours, and their 15 $\mu$m magnitudes. We tried to
select objects with colours typical of pulsating AGB stars with thin
envelopes (Mira-like stars), while avoiding OH/IR stars and other
sources with thick circumstellar envelopes.

Our analysis of the targeted stars' multi-band photometry showed that
these selection criteria were quite reliable.  A comparison between
the DENIS and 2MASS data shows that most of them are variable stars,
and moreover the correlation between the $J$ and \ks\ band brightness
variations is similar to that found in local dust-enshrouded Mira
variable stars \citep{olivier01}.

The IRAS $[12]-[25]$ colours of the SiO targets confirms that they
populate mostly the region IIIa of the van der Veen and Habing
classical two-colour IRAS diagram, which is a region of stars with
moderate mass-loss rates and with silicate feature at 9.7 $\mu$m in
emission. The distribution of the $[12]-[25]$ colours of the SiO
targets overlaps with the distribution of the colours of OH/IR stars,
which however are distributed over a larger and redder range of
colours (mostly in region IIIa and IIIb).  Following the work of
\citet{sevenster02I} and \citet{lumsden02}, those properties can be
translated and seen in the MSX $C-D$ vs.  $A-E$ diagram.  The SiO
targets have a narrower $C-D$ colour range and are located below the
black-body line, differently from the thick-envelope OH/IR stars.

The two subsamples, the MSX-selected objects and the ISOGAL selected
objects have very similar infrared properties but differ slightly in
the average apparent magnitude, the MSX sample being on average a
little brighter. This difference is, however, smaller than the spread
in magnitudes of each subsample. In forthcoming papers we usually will
combine the two subsamples into one.

\begin{acknowledgements}
  We are grateful to G. Simon for providing the DENIS data and to
M. Sevenster for her constructive criticism.  We thank F. Bertoldi,
M. Johnston-Hollitt and F. Schuller for their careful reading and
commenting of an earlier version of the manuscript. \\ We acknowledge
using the cross-correlation package CataPack developed by
P. Montegriffo at the Bologna Observatory.\\ The DENIS project is
supported, in France by the Institut National des Sciences de
l'Univers, the Education Ministry and the Centre National de la
Recherche Scientifique, in Germany by the State of Baden-W\"urtemberg,
in Spain by the DGICYT, in Italy by the Consiglio Nazionale delle
Ricerche, in Austria by the Fonds zur F\"orderung der
wissenschaftlichen Forschung and the Bundesministerium f\"ur
Wissenschaft und Forschung.  The IRAS data base server of the Space
Research Organisation of the Netherlands (SRON).  This publication
makes use of data products from the Two Micron All Sky Survey, which
is a joint project of the University of Massachusetts and the Infrared
Processing and Analysis Center/California Institute of Technology,
funded by the National Aeronautics and Space Administration and the
National Science Foundation.  This research made use of data products
from the Midcourse Space Experiment, the processing of which was
funded by the Ballistic Missile Defence Organization with additional
support from the NASA office of Space Science.  This research has made
use of the SIMBAD data base, operated at CDS, Strasbourg, France.  The
work of MM is funded by the Netherlands Research School for Astronomy
(NOVA) through a {\it netwerk 2, Ph.D. stipend}.
\end{acknowledgements}
    

\end{document}